\documentclass{article}
\usepackage{arxiv}
\usepackage[utf8]{inputenc} 
\usepackage[T1]{fontenc}    
\usepackage{url}            
\usepackage{booktabs}       
\usepackage{amsfonts}       
\usepackage{nicefrac}       
\usepackage{microtype}      
\usepackage{lipsum}		
\usepackage{graphicx}
\usepackage[superscript]{cite}
\usepackage{doi}
\usepackage{graphicx}
\usepackage{setspace}
\usepackage{tocloft}
\usepackage{siunitx}
\usepackage{multirow}
\usepackage{enumitem} 
\usepackage{amsmath,amsfonts,amssymb}
\usepackage{subcaption}
\setlist[itemize]{label=--,nosep}
\usepackage{hyperref}
\usepackage[nameinlink,capitalize]{cleveref}
\title{Generation of biaxially accelerating static Airy light-sheets with 3D-printed freeform micro-optics}

\author{ \href{https://orcid.org/0000-0002-0390-2582}{Yanis Taege}\thanks{\href{mailto:yanis.taege@imtek.uni-freiburg.de?subject=Biaxial paper on arxiv}{yanis.taege@imtek.uni-freiburg.de}}\hspace{4pt}\includegraphics[scale=0.06]{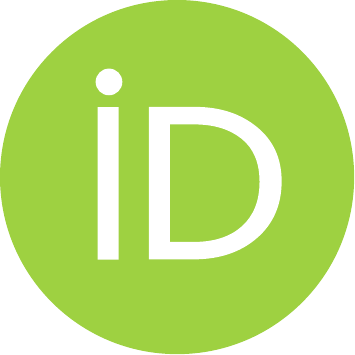} \\
		Laboratory for Micro-optics\\
		Department of Microsystems Engineering (IMTEK)\\
		University of Freiburg, Freiburg, Germany\\
	\And
	Tim Samuel Winter \\
	Laboratory for Micro-optics\\
	Department of Microsystems Engineering  (IMTEK)\\
	University of Freiburg, Freiburg, Germany\\
	\And
	Sophia Laura Schulz \\
	GRINTECH GmbH \\
	Jena, Germany\\
	\And
	\href{https://orcid.org0000-0002-5923-267X}{Bernhard Messerschmidt~\includegraphics[scale=0.06]{orcid.pdf}}  \\
	GRINTECH GmbH \\
	Jena, Germany\\
	\AND
	\href{https://orcid.org/0000-0002-6280-8465}{Çağlar Ataman~\includegraphics[scale=0.06]{orcid.pdf}} \\
	Microsystems for Biomedical Imaging Laboratory\\
	Department of Microsystems Engineering (IMTEK)\\
	University of Freiburg, Freiburg, Germany
}

\renewcommand{\headeright}{Taege et al.}
\renewcommand{\undertitle}{}
\renewcommand{\shorttitle}{Generation of biaxially accelerating static Airy light-sheets}

\newcommand{\FT}{\ensuremath{\mathcal{F}}}
\newcommand{\Ai}{\ensuremath{\text{Ai}}}

\newcommand{\um}{\micro\meter}
\newcommand{\fov}{\ensuremath{\text{FOV}}}
\newcommand{\conv}{\ensuremath{\circledast}}

\hypersetup{
pdftitle={Generation of biaxially accelerating static Airy light-sheets with 3D-printed freeform micro-optics},
pdfsubject={q-bio.NC, q-bio.QM},
pdfauthor={Yanis Taege, Tim Samuel Winter, Sophia Schulz, Bernhard Messerschmidt, Caglar Ataman},
pdfkeywords={light-sheet microscopy, Airy beam, accelerating beams, field curvature, two-photon polymerization},
colorlinks=true, 
allcolors=blue}

\begin{document}
\maketitle

\begin{abstract}
One-dimensional Airy beams allow the generation of thin light-sheets without scanning, simplifying the complex optical arrangements of light-sheet microscopes (LSM) with an extended field-of-view (FOV). 
However, their uniaxial acceleration limits the maximum numerical aperture of the detection objective in order to keep both the active and inactive axes within the depth-of-field. 
This problem is particularly pronounced in miniaturized LSM implementations, such as those for endomicroscopy or multi-photon neural imaging in freely-moving animals using head-mounted miniscopes. 
We propose a new method to generate a static Airy light-sheet with biaxial acceleration, based on a novel phase profile.
This light-sheet has the geometry of a spherical shell whose radius of curvature can be designed to match the field curvature of the micro-objective. 
We present an analytical model for the analysis of the light-sheet parameters and verify it by numerical simulations in the paraxial regime. 
We also discuss a micro-optical experimental implementation combining gradient-index optics with a 3D-nano-printed, fully refractive phase plate.
The results confirm that we are able to match detection curvatures with radii in the 1.5 to 2 mm range. 
\end{abstract}


\clearpage
\section{Introduction}
Light-sheet fluorescence microscopy (LSM) has become an essential imaging modality in the life sciences, providing rapid volumetric imaging with minimal photodamage \cite{Girkin2018}.
Although this makes it an ideal candidate for \textit{in-vivo} imaging such as endomicroscopy\cite{Glover2020,Li2021} and neural imaging in freely moving animals\cite{Flusberg2008,zongLargescaleTwophotonCalcium2022}, translation to miniaturized optical systems has been limited to date\cite{Engelbrecht2010,Sacher2020}.
A critical technology for incorporating LSM into these applications is high performance micro-objectives that can match their macroscopic counterparts in numerical aperture (NA) and field-of-view (FOV)\cite{matzDesignEvaluationNew2016}.
Their performance comes at the cost of field curvature\cite{matzChiponthetipCompactFlexible2017,Pshenay-Severin2021}, which presents an additional challenge for LSM:
With the generally planar geometry of a Gaussian beam, only a limited portion of the illuminated area is brought into perfect focus in the image plane, limiting the effective FOV, imposing limits on the detection NA, or leading to loss in image contrast.
Although the incorporation of scanning micromirrors could address this challenge to some extent\cite{Bakas2021}, it further increases the system complexity and limits the achievable degree of miniaturization.
Using Airy beams \cite{Vettenburg2014} has been proposed to match field curvature due to its uniaxial acceleration when implemented as a static light-sheet.\cite{Niu2017}.
An Airy light-sheet can be generated by imposing a one-dimensional cubic phase profile on a Gaussian beam, which is then focused by a cylindrical lens.
Similar to other configurations using propagation-invariant beams\cite{Fahrbach2010,Kafian2020}, it has been used to excite a larger FOV compared to a Gaussian sheet, while maintaining the same sectioning ability\cite{Nylk2016,Piksarv2017}.
However, incorporating the static Airy light-sheet only partially solves the curvature problem, as the planar Gaussian profile along its inactive axis still introduces a FOV reduction, limiting the performance required for micro-optical applications. 

We have recently shown that this challenge can be addressed by an additional phase modulation on top of the cubic profile needed for the static Airy light-sheet.\cite{Taege2023}
This phase profile depends linearly on the active coordinate and quadratically on the inactive coordinate, and modifies the Gaussian profile of the light-sheet to achieve the same radius of curvature as the Airy beam.
The result is a static, biaxially accelerating light-sheet that is unrestricted in its ability to section, thus increasing the effective FOV of the micro-objective.
In this paper, we provide a detailed analytical derivation of the phase-plate parameterization using typical micro-objective specifications, allowing the light-sheet geometry to be tailored for maximized in-focus excitation.
To verify the validity of our analysis, we perform simulations in the paraxial regime using ray tracing and beam propagation methods.
By designing, implementing, and characterizing a micro-optical excitation system that incorporates a 3D-printed millimeter-scale phase plate, we experimentally demonstrate our ability to generate such a light-sheet in a fully miniaturized fashion.

\section{Theory}
\begin{figure}[p]
	\includegraphics{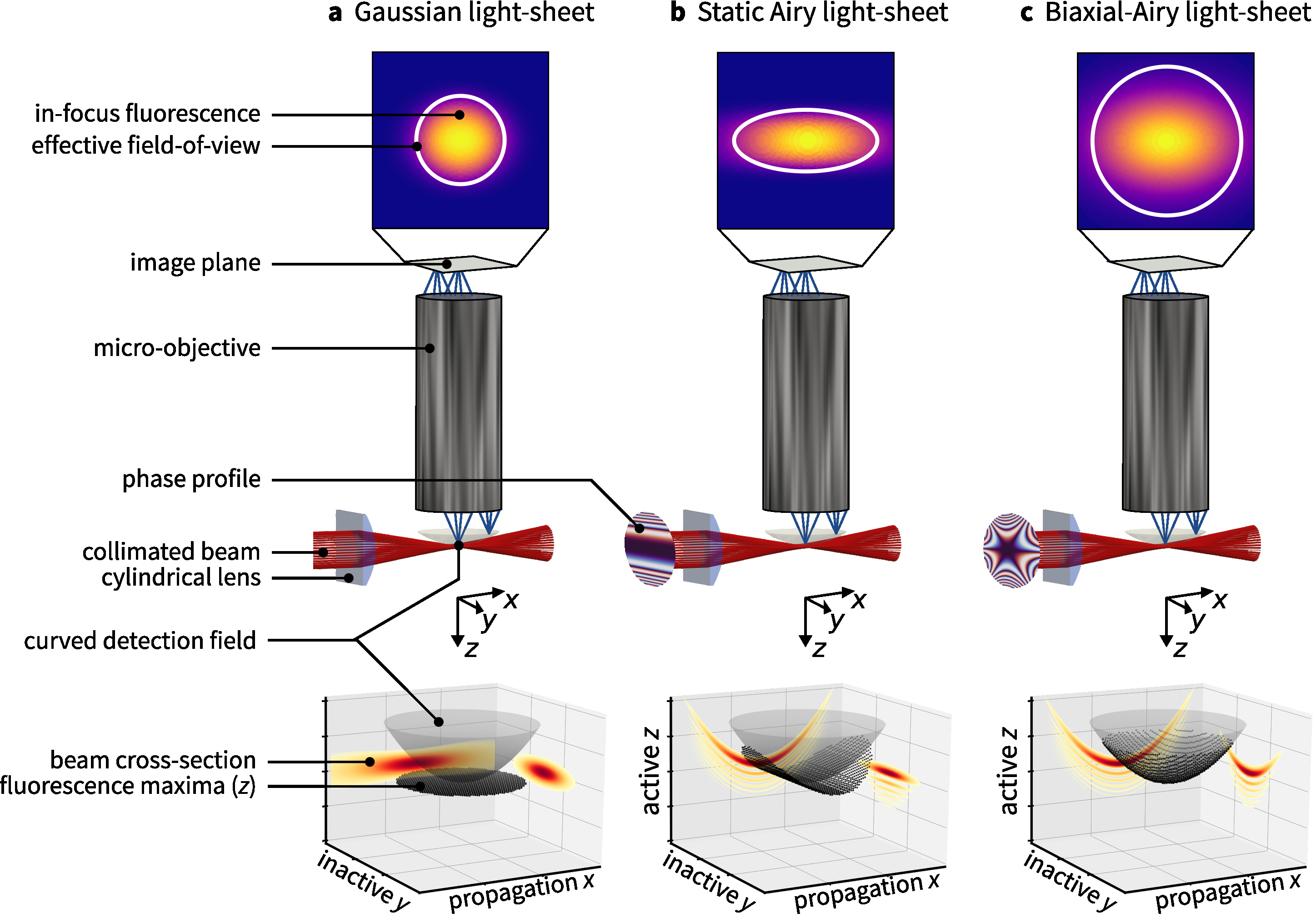}
	\phantomsubcaption\label{fig:introcomparison:a}
	\phantomsubcaption\label{fig:introcomparison:b}
	\phantomsubcaption\label{fig:introcomparison:c}
	\caption[Comparison of different illumination beams for static LSM]{
		Comparison of three different static light-sheet illumination beams for a typical micro-objective for a given light-sheet length and detection numerical aperture.
		The detected in-focus fluorescence, which is imaged by the micro-objective with a curved detection field (gray cone, center) onto the image plane, is plotted on the top.
		In the bottom row, the intensity profiles of the illumination beams are plotted in cross-sections along the  center in the propagation coordinate (XZ-plane, to the back) and the focal plane (XY-plane, to the right). 
		Black dots trace the maximum intensity of the beams along the detection axes, visualizing the mismatch between beam- and detection field-geometry.
		\textbf{a}: Gaussian light-sheet.
		Due to its planar geometry the in-focus fluorescence is significantly smaller than the available FOV.
		\textbf{b}: Static Airy light-sheet.
		Due to the matched curvature along the propagation axis, the effective FOV is increased.	
		In addition, the Airy beam is thinner than the Gaussian beam with the same FOV.
		\textbf{c}: Proposed biaxial Airy beam. By adjusting the curvature also along the inactive direction, the fluorescence collection over the entire illumination area will be in focus.
	} 
\end{figure}

\Cref{fig:introcomparison:a} illustrates the optical architecture of a miniaturized light-sheet microscope with a high NA micro-objective. 
The illumination arm consists of a collimator (not shown) and a cylindrical lens to form a Gaussian light-sheet within the FOV of the micro-objective which is imaging the excited fluorescence onto a flat image sensor.
The field curvature  of the detection objective $z_d$ can be characterized by a radius $r_d \gg \fov$ so that the detection field follows 
\begin{equation}
	z_\text{d}(x,y) \approx \frac{x^2 + y^2}{2r_d}. \label{eq:fieldcurvature}
\end{equation}
As common in LSM we use the coordinate system of the detection objective for the light-sheet, in which 
\begin{itemize}
	\item $z$ is the micro-objective's optical axis, corresponding to the cylindrical lens' active axis,
	\item $x$ is the optical axis of the cylindrical lens corresponding to the beam propagating direction, and
	\item $y$ is the inactive coordinate of the cylindrical lens.
\end{itemize}
\Cref{fig:introcomparison:b} extends the concept to the generation of a static Airy light-sheet, resulting from a one-dimensional cubic phase modulation of the collimated beam incident on the cylindrical lens, which allows field curvature matching in $x$. 
This already increases the effective FOV in this coordinate compared to the Gaussian light-sheet, and improves the sectioning capability due to a thinner light-sheet.
A secondary phase modulation of the proposed phase profile aims at bending the Airy light-sheet along its inactive axis, thus generating a static and yet biaxially accelerating sheet (\cref{fig:introcomparison:c}).
In the next two sections we will discuss the design of the active and inactive phase profiles to achieve biaxial field curvature matching.

\begin{figure}[tp] 
	\begin{center}           
		\includegraphics{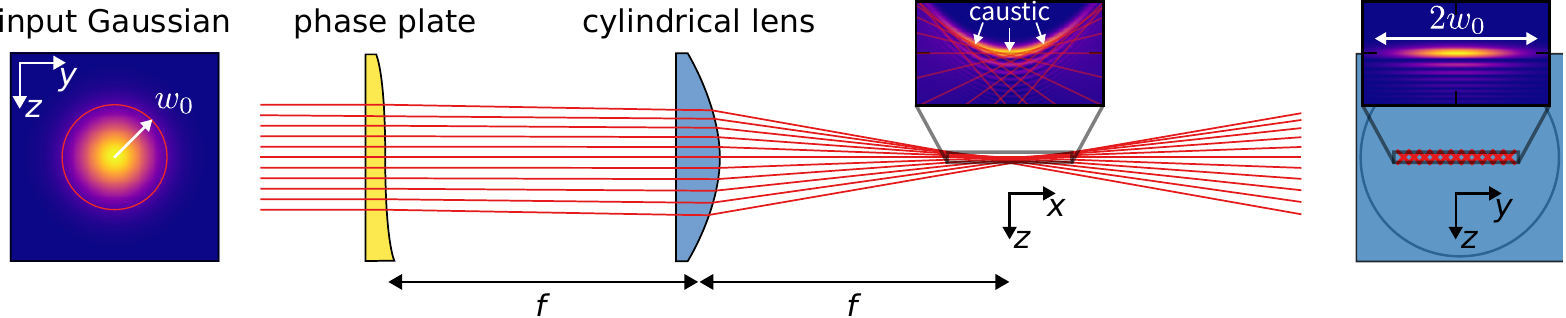}
	\end{center}
	\caption[Schematic of the light-sheet configuration]{Configuration to generate a 1D-Airy light-sheet to illustrate the mathematical derivation.
		An input Gaussian beam with a beam waist of $w_0$ is modulated by a one-dimensional cubic phase plate.
		The phase plate is placed in the back-focal plane (BFP) of a cylindrical lens that focuses the beam in the $z$ coordinate.
		The resulting beam profiles, illustrated by the insets in the XZ and YZ plane, show an Airy profile in the $z$- and a Gaussian profile in the $y$-coordinate.} 
	\label{fig:Airy1D} 
\end{figure}

\subsection{Acceleration along the propagation direction}
An Airy light-sheet can be generated by modulating the collimated Gaussian beam with a beam waist $w_0$ by a one-dimensional cubic phase $\phi_\Ai$ at the back-focal plane (BFP) of the cylindrical lens \cite{Niu2017} along its active direction $z$, such that the electric field follows
\begin{equation}
	E^\text{BFP}_\Ai(y, z) = \exp\left( -\frac{z^2 + y^2}{w_0^2}  \right) \cdot \exp\left[i 2\pi \alpha \left(\frac{z}{w_0}\right)^3\right] \label{eq:eAiry}
\end{equation} 
where $\alpha$ is the amplitude of cubic modulation. 
A schematic depiction of this configuration is shown in \cref{fig:Airy1D}, where the approximate intensity profile around the focal plane can be obtained in the paraxial domain as \cite{Siviloglou2007a,Siviloglou2007b}
\begin{align}
	I_\text{Ai}(x,y,z) \approx \left|\Ai\bigg(  \frac{z}{z_0} -  \frac{x^2}{2r_\Ai z_0} + i \frac{ax}{kz_0^2}\bigg)\right|^2 \cdot \exp \bigg[2a \left(\frac{z}{z_0} - \frac{x^2}{r_\Ai z_0}\right) - \frac{2y^2}{w_0^2}\bigg] \label{eq:iAiry}
\end{align}
where $\Ai(\cdot)$ denotes the Airy function, $x$ is the axial coordinate, $z_0 = f (6\pi\alpha)^\frac{1}{3}/(w_0 k)$ is a characteristic length, $a = \left( 6 \pi \alpha \right)^{-2/3}$ is an apodization factor and $k = 2\pi n/\lambda$ is the wave vector in a medium of refractive index $n$\cite{Taege2022}. 
It has been demonstrated by numerous groups\cite{Siviloglou2007a,Baumgartl2008,Efremidis2019,Chen2011,Hu2012} that this function is highly asymmetrical with prominent side lobes, and its main lobe follows a parabola with a radius of curvature $r_\Ai$ of	
\begin{equation}
	r_\Ai = \frac{6 \lambda n^2 f^3}{w_0^3}\alpha.		\label{eq:rAi}
\end{equation}
Therefore, one must solve \cref{eq:rAi} for $\alpha$ with $r_\Ai = r_d$ to obtain an Airy light-sheet with an acceleration matching the field curvature of the detection objective. 
Using \cref{eq:eAiry} we obtain
\begin{align}
	\phi_\Ai(z) = \frac{2 \pi r_d}{6\lambda n^2 f^3} \cdot z^3.	\label{eq:phiAi}
\end{align}
To finalize the geometry of the light-sheet,the beam needs to be matched to the FOV of the micro-objective, which we define as the area around the focal point within which the intensity remains larger than the half of the maximum value.\cite{Taege2022}
Since \cref{eq:phiAi} does not depend on $w_0$, we can control the geometry of the light-sheet in the propagation coordinate and in the inactive coordinate independently of each other via the focal length and the beam waist.
Explicitly, the FOV along the active direction is obtained from
\begin{equation}
	\fov_x = \frac{w_0 r_d \sqrt{2\ln 2}}{n f},
\end{equation}
and along the inactive direction from
\begin{equation}
	\fov_y = \sqrt{2\ln 2} w_0.
\end{equation}
To obtain a symmetric FOV, the condition $r_d = n f$ must be satisfied, which means that the focal length in the imaging medium must be equal to the radius of curvature.

\subsection{Acceleration along the inactive direction}
To induce a parabolic axial shift of the focal point along the inactive coordinate $y$ as well, we modify \cref{eq:iAiry} by convolving it with a delta distribution  $\delta(\cdot)$ with the required shift in the detection coordinate. 
This operation can be expressed as 
\begin{align}
	I_\text{biaxial}(x, y, z) &= I_\Ai(x,y,z) \conv \delta \left(z - \frac{y^2}{2r_d}\right)  \\
	&\approx \left|\Ai\bigg(  \frac{z}{z_0} -  \frac{x^2 + y^2}{2r_dz_0} + i \frac{ax}{kz_0^2}\bigg)\right|^2 \cdot \exp \bigg[2a \left(\frac{z}{z_0} - \frac{2x^2 + y^2}{2r_d z_0}\right) - \frac{2y^2}{w_0^2}\bigg]  \label{eq:iBiaxial}
\end{align}
where $\conv$ denotes the convolution operation. 
By taking the Fourier transform of the electric field at the focal point $x = 0$ along $z$, we can again obtain the required electric field at the BFP, which can be approximated by 
\begin{align}
	E^\text{BFP}_\text{biaxial}(y, z) \approx\exp \left( -\frac{z^2 + y^2}{w_0^2}   \right)   \cdot  \exp\left[i 2\pi \alpha \left(\frac{z}{w_0}\right)^3\right] \cdot \exp\left(-i\frac{\pi}{r_d\lambda f} \cdot  y^2 z\right). \label{eq:eBiaxial}
\end{align}
Details of this derivation can be found in \cref{sup:pp:derivation}. 
The arguments of the first two exponential functions in \cref{eq:eBiaxial} represent the conventional profile to generate an Airy beam as per \cref{eq:eAiry}, while the last term is an additional, phase-only term that we define as 
\begin{equation}
	\phi_\text{inactive}(y,z) = -\frac{\pi}{r_d\lambda f}\cdot  y^2 z.
\end{equation}
In ray optical terms, this phase profile adds an additional launch angle from the phase plate to the cylindrical lens in its active coordinate, the magnitude of which varies as the square of the inactive coordinate. 
Because of the introduction of this additional angle in the BFP, each one-dimensional Airy ray will be shifted in $z$ at the focal plane of the cylindrical lens.

\subsection{Combined phase profile} \label{sec:ppcombined}
\begin{figure} [ht]
	\begin{center}           
		\includegraphics[width=0.6\textwidth]{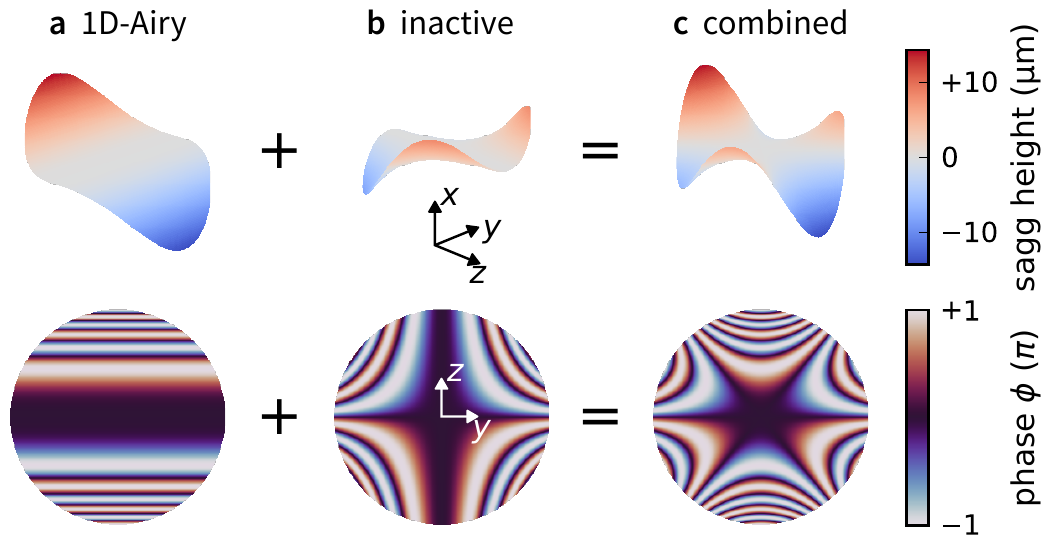}
	\end{center}
	\caption[Combination of the phase profiles]{Schematic illustration of the combined phase plate profiles for a diameter of 1mm. 
		\textbf{a}: Cubic profile $\phi_\Ai$ along the active direction of the cylindrical lens to generate the 1D-Airy beam.
		\textbf{b}: Inactive profile $\phi_\text{inactive}$ that adds a linear phase along the active direction, whose slope depends quadratically on the inactive direction of the cylindrical lens. This profile leads to a focal
		shift in the active coordinate such that it follows the same acceleration as the one induced by the cubic profile. 
		\textbf{c}: Linear combination of the individual profiles that generates the biaxially accelerating light-sheet as given in \cref{eq:hCombined}.} \label{fig:phaseprofiles} 
\end{figure}

A combined phase plate with a sag height of $h_\text{PP}$, which generates both the one-dimensional Airy beam along the propagation coordinate, as well as the focus-shifted Gaussian beam along the inactive coordinate of the beam, can be obtained by
\begin{align}
	h_\text{PP}(y,z) &=  \frac{\lambda \cdot (\phi_\Ai + \phi_\text{inactive})}{2\pi (n_m - 1)} \\
	&= \frac{r_d}{6 n^2 f^3 (n_m-1)}z^3 -\frac{1}{2r_df(n_m-1)} y^2z \label{eq:hCombined}
\end{align}
where $n_m$ is the refractive index of the phase plate. 
The individual profiles are plotted alongside the combined one in \cref{fig:phaseprofiles}. 
An important point about \cref{eq:hCombined} is that, the Airy-related modulation scales linearly with $r_d$, while the modulation related to the inactive profile is inversely proportional to it. 
This imposes a limit on the range of $r_d$ that can be realized:
For lower values of $r_d$, the $\alpha$ parameter is smaller, and the assumptions made in the derivation of the apodized Airy beam become invalid.\cite{Siviloglou2007a, Hu2012}
On the other hand, the modulation generated by the inactive profile dominates the phase plate. 
This leads to larger launch angles, which may violate the paraxial approximation of the optical element.
At the upper end of the limit, it is useful to evaluate whether field curvature remains a problem as the beam becomes flatter and thicker, bypassing the initial problem.

\section{Methods}
\subsection{Simulations}
To assess that curvature-matching in the paraxial regime, we performed ray tracing simulations using a commercial software (OpticStudio 22.1, Zemax LLC). 
The simulation model implemented the configuration shown in \cref{fig:Airy1D} with the phase profile described in \cref{sec:ppcombined}.
We calculated the curvature of the beam in the active direction by evaluating the caustic\cite{Wen2017} of 51 rays displaced in the active $z$-coordinate (XZ-section in \cref{fig:Airy1D}).
For the inactive direction, we obtained the the displacement of the rays in the image plane by tracing 51 rays that were initially displaced along the $y$-direction.
To further investigate the validity of \cref{eq:iBiaxial}, we propagated the field defined in \cref{eq:eBiaxial} using a two-dimensional FFT-based beam propagation method (BPM) in the Fresnel regime \cite{data_underpinning}.
Propagation was performed to match the ray tracing simulations, which include the same beam waist, phase plate profile, and focal length of the cylindrical lens.
3-dimensional stacks were calculated within a range of $\pm\SI{300}{\um}$ around the focal point with a step size of $\Delta z = \SI{5}{\um}$, and a grid size of  $(n_x \times n_y) = (2^{12} \times 2^{12})$ pixels with a width of $\Delta x = \Delta y = \SI{0.2}{\um}$. 
We set the input parameters with an excitation wavelength $\lambda = \SI{850}{nm}$, an input beam waist $w_0 = \SI{240}{\um}$, an effective focal length of the cylindrical lens $f_z = \SI{1.4}{mm}$ and a refractive index of the phase plate $n_m =1.509$\cite{Schmid2019} to match the experimental setup described below.
Two radii of curvature were chosen in a typical range for micro-objectives of \SI{1.5}{mm} and \SI{2}{mm}. \cite{matzDesignEvaluationNew2016,matzChiponthetipCompactFlexible2017,Schulz2023}

\subsection{Fabrication and characterization}
Phase plates with a diameter of \SI{0.78}{mm} were fabricated using a commercial two-photon polymerization based 3D printer (PPGT+, Nanoscribe GmbH \& Co KG) with a commercial resin (IPS, NanoScribe GmbH \& Co KG). 
The height profiles were pre-processed and subsequently fabricated on ITO-coated glass substrates using an in-house developed slicing and printing strategy previously described \cite{Taege2022b}.
After development, the height profiles of the phase plates were measured by white light interferometry (WLI; NewView 9000, Zygo Corporation).
A 0.3-NA objective (I100384, Zygo Cooperation) was used to obtain the surface profile of the entire plate and then fitted with standard Zernike polynomials to quantify the shape deviation.
Since the tilt of the printing field in this system can only be controlled to within a few microns from print to print, the tilt Zernike coefficients were not included in the deviation analysis.
The surface roughness was estimated by measuring a $\SI{25}{\um}\times\SI{25}{\um}$ area of the phase plate with a 0.8-NA objective (I190053, Zygo Cooperation). A high-pass Gaussian spline filter with a cutoff frequency of $\SI{25}{\um}$ was applied and the resulting root mean square error (RMSE) was calculated. \cite{ISO25178}

\subsection{Experimental setup and analysis of the beam profiles}
\begin{figure}[thb]
	\begin{center}           
		\includegraphics{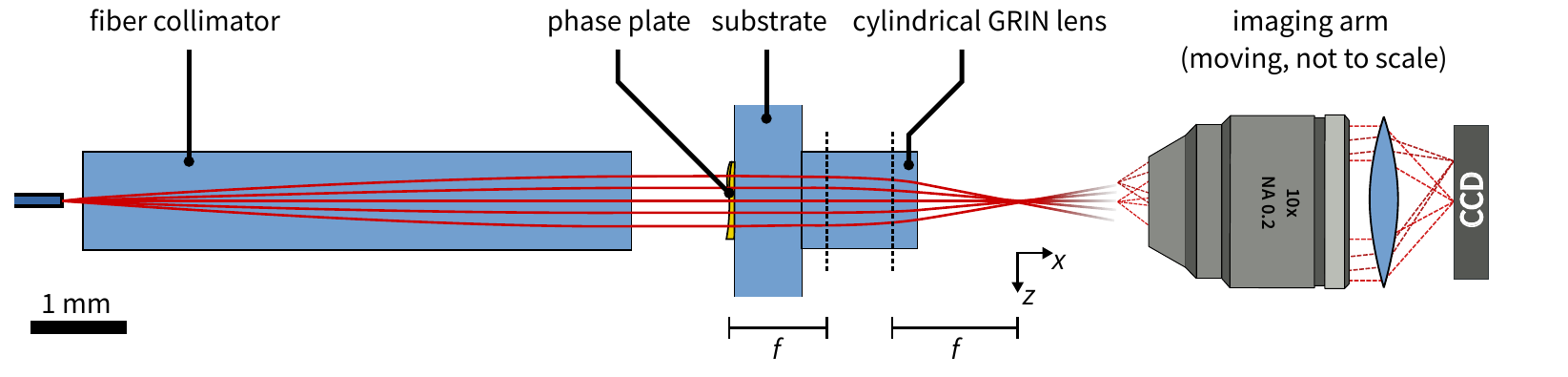}
	\end{center}
	\caption[Experimental setup]{
		Experimental setup used to measure the biaxially accelerating beam profiles following the configuration in \cref{fig:Airy1D}.
		For a micro-optical implementation, input beam shaping was performed by a fiber-based collimation unit.
		The glass substrate on which the phase plate has been fabricated features a cylindrical GRIN lens at its back which focuses the modulated beam from its backfocal plane to the focal plane.
		Around this plane the light-sheet is imaged by moving an imaging arm, consisting of a 0.2-NA detection objective, a relay lens and a CCD, in the propagation direction of the beam.
	} \label{fig:experimental} 
\end{figure}

A cylindrical gradient-index (GRIN) lens with an effective focal length of \SI{1.4}{mm} (Grintech GmbH) was bonded to the back of the substrate after fabrication so that the phase plate was located at its back focal plane. 
The entire assembly was illuminated with a fiber-based collimator (Grintech GmbH) using a SLED with a wavelength of \SI{850}{nm} as the light source (\cref{fig:experimental}).
The beam waist of the illumination beam after the collimator was $w_0 = \SI{240}{\micro\meter}$.
In order to obtain the 3D-dimensional beam profile in the vicinity of the focal plane of the cylindrical lens, the beam was imaged onto a CCD (UI-1240SE-NIR, IDS GmbH) using a 0.28-NA objective (Plan Apo 10x, Mitutoyo AC) and a relay lens, while the entire system was translated along the propagation direction of the light-sheet.
To compare the analytical analysis, simulations and experiments, two-dimensional cross sections of the beam intensity along its propagation direction at ($y = 0$) and along the inactive direction at the focal plane ($x = 0$) were measured.
The thickness $\Delta z$ and the $\fov_{x,y}$ were determined by finding the full-width at half-maximum of the beam intensity in the respective direction.
The radius of curvature in $x$ and $y$ was determined by finding the location of the maximum intensity in the respective coordinate and fitting \cref{eq:fieldcurvature} to these data points using a least-squares algorithm.\cite{scipy}

\clearpage
\section{Results}
\begin{figure}[h]
	\begin{center}           
		\includegraphics{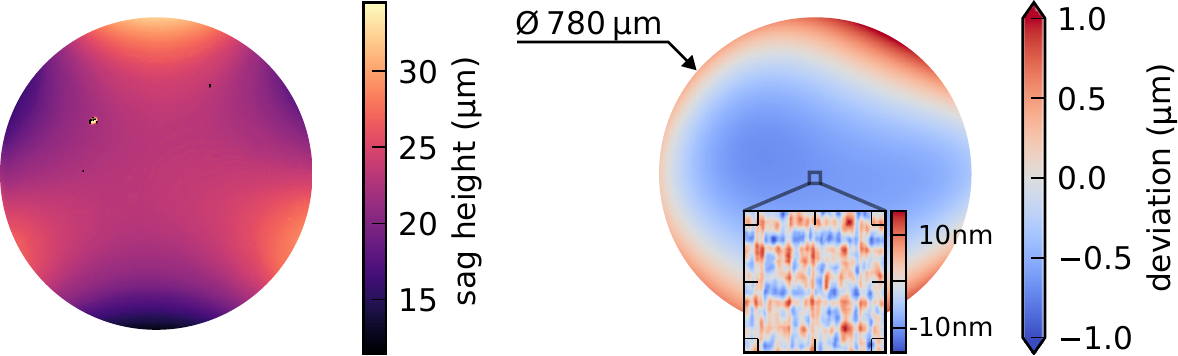}
	\end{center}
	\caption[Characterization of the phase plate]{
		Characterization of the phase plate profile with $r_d = \SI{2}{mm}$ using white-light-interferometry.
		The left image depicts the measured surface profile, and the right is the tilt-corrected deviation from the ideal surface i.
		The inset on the right provides the  high-pass filtered, high-NA scan for roughness estimation.
	} \label{fig:pp:experimental} 
\end{figure}

\cref{fig:pp:experimental} shows the characterization of the surface profile obtained with the WLI.
The surface errors become more pronounced towards the edge, where lift-off from the substrate is expected, but were found to be less than \SI{1}{\um} over a diameter of \SI{780}{\um}. 
With an RMS surface roughness below \SI{10}{nm}, the phase plate can be considered an optical quality surface.

\begin{figure}[hp]
	\begin{center}           
		\includegraphics[width=\textwidth]{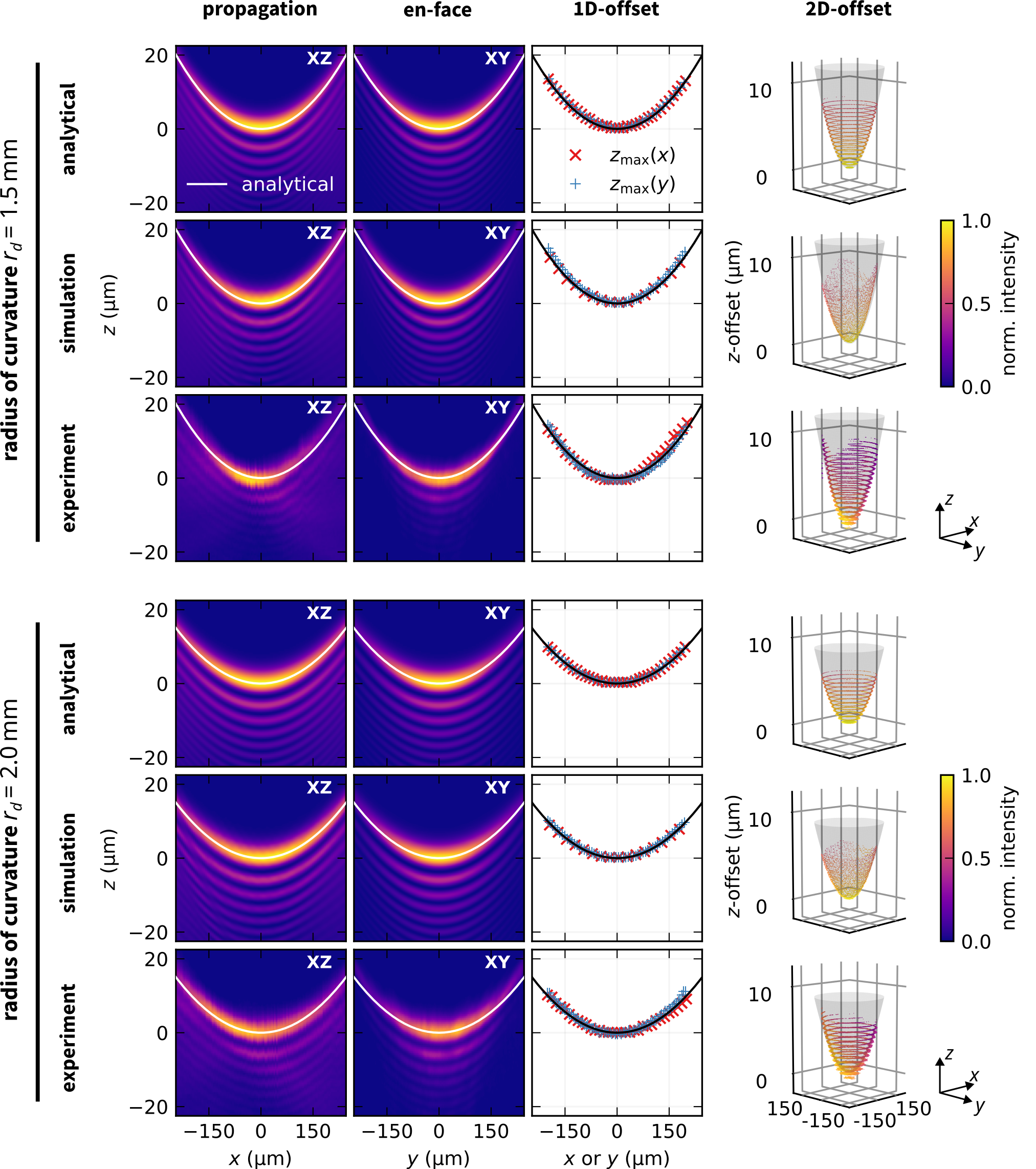}
	\end{center}
	\caption[Simulation and experimental results]{Biaxially accelerating beam profiles obtained by the analytical evaluation (\cref{eq:iBiaxial}), the BPM-simulations and the beam profiling experiments.
		The central cross-section of the beam in propagation direction (XZ-plane) and along the inactive coordinate (XY-plane) are plotted and show the typical Airy-profile of the beam, though with an acceleration also along the latter coordinate.
		The $z$-offsets of the intensity maxima are plotted in $x$ and $y$ of both one-dimensional cross-sections.
		To also show that the main lobe's intensity resembles a spherical cap, the $z$-offsets are plotted along both dimensions simultaneously and color-coded with the respective intensity. 
		The evolution of the design radii are shown alongside each plot.        
	} \label{fig:results} 
\end{figure}

The ray tracing simulations yielded radii of curvature that matched the design radii within numerical accuracy.
Furthermore, the simulated beam profiles shown in \cref{fig:results} agree with those generated by evaluating \cref{eq:iBiaxial}.
Deviations can be narrowed down to an asymmetric intensity distribution around the focus below \SI{10}{\um}, and predicted offsets in $z$ are consistent with both design and numerical predictions.
The radii of curvature obtained from measured data matched the design values in most cases:
For $r_d = \SI{1.5}{mm}$, the radii in the propagation coordinate and in the inactive coordinate were found to be \SI{0.2(1)}{mm} and \SI{0.1(1)}{mm} smaller than the design value, respectively.
For $r_d = \SI{2.0}{mm}$, the radius of the active coordinate is equal to the designed radius, while the inactive coordinate is \SI{0.2(1)}{mm} smaller than the designed radius.
The beam profiles themselves show the typical Airy profile, although a reduced side lobe intensity is observed with an underestimation of the FOV in propagation direction of $20 - \SI{36}{\percent}$ and $5 - \SI{10}{\percent}$ in the inactive coordinate.

\begin{table}[h]
	\caption[Resulting radii of curvature]{
		Comparison of the radius of curvature determined by \cref{eq:iBiaxial}, ray-tracing and BPM simulations with the experimental measurements.
		The corresponding beam profiles of the latter three are also shown in \cref{fig:results}. 
		Since the differences between the analytical and ray-tracing results are within the limits of numerical accuracy, they are not shown explicitly. 
		\vspace{5pt}
	} \label{tab:results}
	\centering
	\begin{tabular}{llrr}
	\toprule 
	design radius						&							& active coordinate & inactive coordinate \\ \midrule 	
	\multirow{4}{*}{$r_d = \SI{1.5}{mm}$} 	& analytical				& \SI{1.5}{mm}      & \SI{1.5}{mm} \\
	& simulation (ray-tracing) 	& \SI{1.5}{mm}  	& \SI{1.5}{mm} \\
	& simulation (BPM)  		& \SI{1.5(1)}{mm}  	& \SI{1.4(1)}{mm} \\
	& experiment 				& \SI{1.3(1)}{mm} 	& \SI{1.4(1)}{mm} \\ \cmidrule(lr){2-4}
	\multirow{4}{*}{$r_d = \SI{2.0}{mm}$} 	& analytical				& \SI{2.0}{mm}      & \SI{2.0}{mm} \\
	& simulation (ray-tracing) 	& \SI{2.0}{mm}  	& \SI{2.0}{mm} \\
	& simulation (BPM)  		& \SI{2.1(1)}{mm}  	& \SI{2.0(1)}{mm} \\
	& experiment 				& \SI{2.0(1)}{mm} 	& \SI{1.8(1)}{mm} \\ \bottomrule	
\end{tabular}
\end{table}

\clearpage
\section{Discussion and conclusions} \label{sec:discussion}
The excellent agreement between the results of our analytical treatment and the ray traced simulations is a good indication of the validity of the design approach within the paraxial regime. 
Limitations of this approximation may arise in different scenarios:
As discussed in \cref{sec:ppcombined}, lower values of $r_d$ will lead to a reduced modulation of the Airy profile.
Therefore, we expect the radius to be underestimated in the propagation coordinate, as observed in the experimental data.
In the inactive direction, we also expect to see an underestimated radius, but for a different reason: 
The higher additional deflection leads to a gradual violation of the paraxial approximation, resulting in aberrations and thus lower radii.
On the other hand, we observed the Airy radius to be the same as the design radius for larger values of $r_d$, while the radius along the inactive direction is slightly smaller.
We attribute this to a manufacturing effect as for smaller values of $r_d$, the sag height decreases and therefore requires more accurate manufacturing capabilities.
Both effects can lead to the deviations reflected in the experimental results.
Nevertheless, we have been able to adjust both radii to a degree that does not affect the improvement in effective FOV, which was the underlying design goal.

In summary, we have derived a phase profile to generate an Airy light-sheet that follows the curved image plane of a typical micro-objective, while incorporating the advantages of using an Airy beam to provide a larger FOV and better sectioning. 
Ray tracing and BPM simulations in the paraxial regime validated the analytical analysis, with deviations between the two within numerical accuracy. 
A micro-optical implementation with mm-scale 3D-printed refractive phase plate was developed and combined with GRIN optics, and experimentally demonstrated to achieve curvature matching in the design range.

\section*{Acknowledgements}
We express our gratitude to Anders Kragh Hansen (Technical University of Denmark) for having the initial idea of incorporating a quadratic phase profile for curvature, as well as to Anja Borre, Madhu Veettikazhy and Peter Andersen for discussions on the phase distribution.
This project has received funding from the European Union's Horizon 2020 research and innovation program under grant agreement No 871212.

\section*{Data availability}
The simulation and evaluation code used to generate the results presented is available upon request.

\section*{Disclosures}
Sophia Laura Schulz and Bernhard Messerschmidt are full-time employees of the company GRINTECH GmbH.

\bibliographystyle{spiejour}
\bibliography{bibliography}   

\listoffigures
\listoftables

\appendix 
\newpage
%

\section{Derivation of the phase profile} \label{sup:pp:derivation}
To obtain the required phase profile, we need an expression of the biaxial electric field at the focal plane $x=0$, which we can approximate from the apodized Airy beam convoluted by the delta distribution $\delta$ as in \cref{eq:iBiaxial} such that
\begin{equation}
E_\text{biaxial}(x=0, y, z) = \Ai\left(\frac{z}{z_0}\right) \cdot \exp \left(\frac{az}{z_0} - \frac{y^2}{w_0^2}\right) \conv \delta\left(z - \frac{y^2}{2r_d}\right). 
\end{equation}
The electric field at the backfocal plane can be obtained by taking the Fourier transform along the active direction of the lens $\FT_z\{\cdot\}$, i.e.
\begin{align}
E^\text{BFP}_\text{biaxial}(y, z) &= \FT_z \left\{ \Ai\left(\frac{z}{z_0}\right) \cdot \exp \left(\frac{az}{z_0} - \frac{y^2}{w_0^2}\right) \conv \delta\left(z - \frac{y^2}{2r_d}\right) \right\} 
\end{align}
Since the Gaussian component is only dependent on $y$, it is neither affected by $\FT_z$, nor by the delta distribution and can thus be separated.
\begin{align}
E^\text{BFP}_\text{biaxial}(y, z) &= \FT_z \left\{ \Ai\left(\frac{z}{z_0}\right) \cdot \exp \left(\frac{az}{z_0}\right) \conv \delta\left(z - \frac{y^2}{2r_d}\right) \right\} \cdot \exp\left(- \frac{y^2}{w_0^2}\right). \label{eq:eBiaxialBFP:inter}
\end{align}
Using the convolution theorem we can also simplify the convolution operation to a product, such that $\FT_z$ operates individually on the apodized Airy function, whose identity is known from \cref{eq:eAiry}. 
The Fourier transform of a linear delta distribution is an exponential function 
\begin{align}
\FT_z\left\{\delta\left(z - \frac{y^2}{2r_d}\right) \right\} \propto \exp\left(-i\frac{k_z y^2}{2r_d}\right)
\end{align}
where $k_z \approx 2\pi z/(\lambda f)$ is the wavevector in the active coordinate.
Inserting those identities into \cref{eq:eBiaxialBFP:inter} leaves
\begin{align}
E^\text{BFP}_\text{biaxial}(y, z) &= \FT_z \left\{ \Ai\left(\frac{z}{z_0}\right) \cdot \exp \left(\frac{az}{z_0}\right)\right\} \cdot \FT_z \left\{\delta\left(z - \frac{y^2}{2r_d}\right) \right\} \cdot \exp\left(- \frac{y^2}{w_0^2}\right)\\
&\approx \exp\left(-\frac{z^2}{w_0^2}   \right)   \cdot  \exp\left[i 2\pi \alpha \left(\frac{z}{w_0}\right)^3\right] \cdot \exp\left(-i\frac{\pi}{r_d\lambda f} \cdot  y^2 z\right) \cdot \exp\left( \frac{-y^2}{w_0^2}\right) \\
&= \exp\left(-\frac{z^2 + y^2}{w_0^2}\right) \cdot \exp\left[i 2\pi \alpha \left(\frac{z}{w_0}\right)^3\right] \cdot \exp\left(-i\frac{\pi}{r_d\lambda f} \cdot y^2z\right) \label{eq:eBiaxial:derived}
\end{align}
which is a Gaussian beam (first exponential function), on which the cubic Airy phase (second exponential argument), as well as the biaxial tilt (third exponential argument) is imposed. $\blacksquare$

\subsection{Approximations}
We would like to underline the fact that several assumptions were made in the derivation of \cref{eq:eBiaxial:derived} which can be summarized by the following points
\begin{itemize}
\item The Rayleigh range of the input Gaussian beam is much larger than the propagation distance from the backfocal to the focal plane of the lens, i.e. $x_r \gg 2f$. 
This ensures that the light-sheet does not expand in $y$.
\item All approximations made are strictly in the paraxial regime, such that the Abbe sine condition is fulfilled, i.e. $k_z/k \approx z/f$.
\item There are several approximations made for the expression of the Airy beam, of which one is that the apodization $a$ must be sufficiently strong, i.e. $a \ll 1$. 
For our application, this condidition is reasonably fulfilled if $\alpha > 2$.
\item All calculations are strictly proportionalities (e.g. constants like the impedance or scaling factors like $\sqrt{2\pi}$ are not mentioned explicitly).
\end{itemize}
Any violation of those assumptions may result in aberrations of the Airy beam, which lead to a variation in length and thickness of the light-sheet, and in mismatch of the field curvature as discussed in \cref{sec:discussion}.

\section{Other inactive geometries} \label{sup:slmGeometries}
In general, the argument of the $\delta$ function in \cref{eq:eBiaxial:derived} can be modified to resemble an arbitrary focal shift in $z$ depending on a function $g(y)$, i.e.
\begin{equation}
s(z, y) = \delta (z - g(y)).
\end{equation}
An example for a benchtop application is the combination of a spatial light modulator (SLM) with a cylindrical lens to generate a 1D-Airy light-sheet.
If the imaging arm is rolled with respect to the detection arm the light-sheet will intersect the focal plane at an angle.
By setting $g(y)$ to match this tilt one can compensate this misalignment without moving any mechanical components by  employing a phase of
\begin{equation}
\phi_\text{inactive} = k_z \cdot g(y)
\end{equation}
onto the SLM.

\section{Further detailed results}

\begin{table}[hp]
\caption[Detailed results]{Detailed results accompying those in \cref{tab:results}.
	In addition to the radii of curvature, the $\fov_{x,y}$ in each direction is given, as well as the thickness of the light-sheet $\Delta z$.}\label{tab:results:detailed}
\centering
\begin{tabular}{llrrrrr}
	\toprule
	& & \multicolumn{2}{c}{active $x$}         & \multicolumn{2}{c}{active $y$}            & \\ \cmidrule(lr){3-4} \cmidrule(lr){5-6}
	design radius        & & $r_x$ (\si{mm}) & $\fov_x$ (\si{\um})  & $r_y$ (\si{mm}) & $\fov_y$ (\si{\um})   & $\Delta z$ (\si{\um}) \\ \midrule
	\multirow{4}{*}{$r_d =\SI{1.5}{mm}$} & analytical &1.5 & 354 & 1.5 & 299 & 4.0\\
	& ray-traced & 1.5 & - & 1.5 & - & -\\
	& BPM &\SI{1.5(1)}{} & 362 & \SI{1.4(1)}{} & 301 & 4.0\\
	& experiment &\SI{1.3(1)}{} & 216 & \SI{1.4(1)}{} & 291 & 3.7\\ \cmidrule(lr){2-7}
	\multirow{4}{*}{$r_d =\SI{2.0}{mm}$} & analytical &2.0 & 462 & 2.0 & 301 & 4.4\\
	& ray-traced & 1.5 & - & 1.5 & - & -\\
	& BPM &\SI{2.1(1)}{} & 482 & \SI{2.0(1)}{} & 301 & 4.4\\
	& experiment &\SI{2.0(1)}{} & 361 & \SI{1.8(1)}{} & 270 & 3.9\\
	\bottomrule
\end{tabular}
\end{table}

\end{document}